\documentclass[12pt]{article}
\usepackage{cite}
\usepackage{verbatim}
\usepackage{amsmath}
\usepackage{amssymb}
\usepackage{amsfonts}
\usepackage{latexsym}
\usepackage{epsfig}
\usepackage{color}

\def\bea{\begin{eqnarray}}
\def\eea{\end{eqnarray}}
\def\beq{\begin{equation}}
\def\eeq{\end{equation}}

\def\a{& \hspace{-7pt}}

\def\al{{\alpha^\prime}}

\begin{document}

\thispagestyle{empty}
\vspace*{.5cm}
\noindent
DESY 04-214 \hspace*{\fill} November 12, 2004\\
HD-THEP-04-47

\vspace*{1.6cm}

\begin{center}
{\Large\bf Gauge Unification\\[0.3cm]
in Highly Anisotropic String Compactifications}
\\[2.3cm]
{\large A. Hebecker$\,^a$ and M. Trapletti$\,^b$}\\[.5cm]
{\it $^a$Institut f\"ur Theoretische Physik, Universit\"at Heidelberg,
Philosophenweg 16 und 19, D-69120 Heidelberg,
Germany}\\[.2cm]
{\it $^b$Deutsches Elektronen-Synchrotron, Notkestra\ss e 85, D-22603 Hamburg,
Germany}
\\[.4cm]
{\small\tt (\,hebecker@thphys.uni-heidelberg.de\,,\,\,
michele.trapletti@desy.de\,)}
\\[1.3cm]

{\bf Abstract}\end{center}

\noindent
It is well-known that heterotic string compactifications have, in spite of 
their conceptual simplicity and aesthetic appeal, a serious problem with 
precision gauge coupling unification in the perturbative regime of 
string theory. Using both a duality-based and a field-theoretic definition of 
the boundary of the perturbative regime, we reevaluate the situation in a 
quantitative manner. We 
conclude that the simplest and most promising situations are those where 
some of the compactification radii are exceptionally large, corresponding to 
highly anisotropic orbifold models. Thus, one is led to consider 
constructions which are known to the effective field-theorist as
higher-dimensional or orbifold grand unified theories (orbifold GUTs). In 
particular, if the discrete symmetry used to break the GUT group acts freely, 
a non-local breaking in the larger compact dimensions can be realized, 
leading to a precise gauge coupling unification as expected on the basis of 
the MSSM particle spectrum. Furthermore, a somewhat more model dependent but 
nevertheless very promising scenario arises if the GUT breaking is restricted 
to certain singular points within the manifold spanned by the larger 
compactification radii.

\newpage

\section{Introduction}
Calabi-Yau or orbifold compactifications of the heterotic string~\cite{
Gross:1985rr} in its perturbative regime provide one of the most direct and 
beautiful paths from string theory to supersymmetric particle 
phenomenology~\cite{Candelas:1985en,Dixon:1985jw}. However, at the 
quantitative level, they are known to have serious problems related to the 
phenomenologically preferred not-so-small value of the unified gauge 
coupling and the values of Planck scale and Grand Unification 
scale~\cite{Kaplunovsky:1985yy}. Such difficulties may be avoided by 
appealing to threshold corrections reconciling $M_{GUT}$ with 
$M_{string}$~\cite{Dixon:1990pc,Nilles:1997vk} (see also 
\cite{Gaillard:1992yb,Faraggi:1993sr,Dienes:1995sq} and, very recently, 
\cite{Ross:2004mi}). The problems may also be resolved in a 
regime where either the space-time or the world sheet expansion parameter of 
the string, or both, are not small~\cite{Witten:1996mz,Banks:1996ss, 
Caceres:1996is}. However, the striking feature of quantitative gauge 
coupling unification in the minimal supersymmetric standard model (MSSM), 
which strongly supports the ideas of low-scale supersymmetry and thereby, 
indirectly, of the superstring, is in general lost. 

In the present paper we approach the above set of problems focussing on 
new energy scales introduced in the compactification procedure. We analyse 
the issue of gauge coupling unification in heterotic orbifold models and 
conclude that the most promising constructions are those where the compact 
space is highly anisotropic, i.e. one or two of the 6 compactification radii 
are exceptionally large~\cite{Witten:1996mz}. In other words, taking the 
empirical fact of precision gauge unification in the MSSM seriously 
(see~\cite{Amaldi:1991cn} and refs. therein), one is forced 
into a regime that corresponds, from the low-energy perspective, to 
higher-dimensional grand unified theories (orbifold GUTs)~\cite{
Kawamura:2000ev,Hall:2001pg,Hebecker:2001wq,Asaka:2001eh}. Possibilities for 
realizing such GUT models in explicit heterotic string constructions have 
recently been explored in~\cite{Kobayashi:2004ud,Forste:2004ie, 
Kobayashi:2004ya} (for an earlier brief discussion see~\cite{Hall:2001xb}). 

To be more specific, we identify at least two interesting options for 
realizing precision gauge coupling unification. In both cases, this can be 
done in spite of possible non-perturbative stringy effects, which become 
relevant at the scale of the small compactification radii. In the first,
conceptually simpler case, the geometry of the manifold defined by the large 
compactification radii is used to realize a non-local breaking of the GUT 
group. The corresponding compactification scale is the GUT scale, above which 
one enters a higher-dimensional field theory regime with a locally unbroken 
GUT symmetry. Any, possibly strongly coupled, string physics at or near the 
scale set by the small compactification radii does not influence gauge 
coupling unification which is, in fact, calculable within higher-dimensional 
field theory. The crucial technical point is that the GUT breaking has to be 
realized using a freely acting orbifold (see~\cite{Vafa:1995gm} for 
applications to non-local SUSY breaking and~\cite{Hebecker:2003we} for a 
recent field theoretic example with gauge symmetry breaking). A possible 
disadvantage of this scenario is that it may be difficult to find realistic 
models within this fairly constrained class of orbifold constructions. We 
consider this an interesting challenge for the future. 

In the second, more familiar case, the GUT group is broken only
at certain points within the space spanned by the large compactification 
radii. In spite of this local breaking, gauge unification at the level of 
one-loop precision can be maintained.
This is the argument of ``bulk dominance'' familiar in the context 
of orbifold GUTs. From a stringy perspective this means that, even though 
the actual unification takes place in a regime that is hard to control 
quantitatively (small compactification radii near the string scale and/or 
large string coupling), the resulting uncalculable gauge coupling corrections 
are volume suppressed. Within this large volume (bulk) the GUT group is 
unbroken and strong coupling effects do not affect standard model (SM) gauge 
coupling differences. Such a scenario has the disadvantage that, in general, 
the logarithmic running of coupling differences continues above the 
compactification scale corresponding to the large radii~\cite{Nomura:2001mf,
Hall:2001pg,Hebecker:2001wq}. This modified running depends on the details of 
the specific model and may significantly affect the prediction of low-energy 
gauge couplings. Thus, even though very promising concrete realizations are 
known in the field-theory context, extra assumptions beyond the simple use 
of the MSSM particle spectrum enter the analysis.

The paper is organized as follows. In Sect.~\ref{wc} we attempt to make 
known arguments specifying the boundary of the weak-coupling regime of 10d 
heterotic string theory as quantitative as possible. This can be done by 
appealing to duality and identifying the point when massive states of the 
dual type I string or heterotic $M$ theory become lighter than the lowest 
massive heterotic string level. Alternatively, the same result is obtained 
by identifying the loop expansion parameter of the effective 10d gauge 
theory with the cutoff specified by the lowest-lying massive string state 
and requiring this parameter to be smaller than one. Writing the familiar 
${\cal N}\!=\!1$ $d\!=\!10$ supergravity action at fixed dilaton VEV as
\beq
S=-\int d^{10}x\frac{\sqrt{G}}{(2\pi)^7g^2}\left(\frac{32}{\al^4} 
R+\frac{4}{\al^3} {\rm Tr_V} F_{\mu\nu}F^{\mu\nu}\right)\,,\label{fe}
\eeq
we find that the dimensionless loop expansion parameter is $g^2$ and the 
theory is weakly coupled for $g<1$. Notice the presence of a large factor 
$(2\pi)^7$ in front of $g^2$. (The dilaton dependence is recovered by 
recalling that $g\sim\exp(\phi)$.)

Using this result and a network of dualities~\cite{Antoniadis:1999rm}, we 
discuss in Sect.~\ref{ia} 
the phenomenology of compactifications. We begin by quantifying the known 
difficulty of obtaining a realistic GUT from an isotropic heterotic string 
compactification. We then analyse the situation in anisotropic 
compactifications with one or two of the radii exceptionally large. 
Particular attention is paid to the question of how small the smaller radii 
can be made before perturbative control is completely lost. We emphasize the 
implications of T duality with Wilson lines and comment on the consequences 
of ${\cal N}\!=\!4$ or ${\cal N}\!=\!2$ supersymmetry in the 5 or 6d bulk. 

We find that, for heterotic SO(32) string theory, a compactification with $d 
\!=\!2$ larger extra dimensions with inverse length set to the MSSM GUT scale 
$2\times 10^{16}$ GeV is possible. This implies the viability of a non-local 
GUT breaking scenario with correct breaking scale. In the $d\!=\!1$ case, 
such a non-local breaking scenario can not be realized. However, we find it 
possible to obtain a significant separation of 5d compactification scale 
and string scale allowing for an analysis in the framework of 5d orbifold 
GUTs. Similar studies are carried out in the $E_8\times E_8$ case, but the 
lack of $T$-duality on the $M$-theory side limits our power to control 
perturbativity.\footnote{
In the special case of Wilson line breaking to SO(16)$\times$SO(16), the 
E$_8\times$E$_8$ theory is dual to the SO(32) theory and the stronger SO(32) 
results apply~\cite{Ginsparg:1986bx}.
}
Using conservative estimates we find that, both in the 
$d\!=\!2$ and $d\!=\!1$ case, the large $M$-theory radius puts us in a 
phenomenologically more difficult situation than in the SO(32) case. Thus, 
heterotic SO(32) scenarios, which were neglected in the past and have only 
recently been subject to the attempt of a complete 
classification~\cite{Choi:2004wn}, may be favoured. 

In Sect.~\ref{nln}, we describe explicitly how the various allowed 
compactification scales discussed above from a more formal perspective 
constrain the construction of realistic string GUTs. This analysis, which 
determines the main physical conclusions of the paper, reveals two main 
phenomenological options. These two options, both of which have already 
been briefly outlined above, correspond to non-local breaking in two extra 
dimensions (with radius set by the inverse GUT scale) and to local 
(fixed-point) breaking at singularities in one or two larger compact 
dimensions. Sect.~\ref{nln1} explains how to realize non-local GUT breaking 
in a heterotic orbifold construction and Sect.~\ref{locbrk} provides a 
detailed analysis of mass scales in 5d models with local GUT breaking and 
large compactification radius (string-based orbifold GUTs). Amusingly, for 
$g\!=\!1$ and with the small radii set to the inverse string theory 
threshold, one finds a compactification scale $M_c\simeq 3.4\times 
10^{16}\,{\rm GeV}$. (We are presently unable to make use of this intriguing 
proximity to the GUT scale in a realistic model). 

Section~\ref{ex} provides some illustrative concrete examples for the above 
scenarios. In particular, in Sect.~\ref{ex1} a simple $Z_2\times Z_2'$ 
heterotic orbifold model is proposed which contains, at an intermediate 
energy scale, a 6d SO(10) gauge theory broken non-locally to a 4d Pati-Salam 
model using a freely acting discrete symmetry. Furthermore, 
Sect.~\ref{OGUT5D} relates our discussion of models with local breaking to 
specific 5d orbifold GUTs and recent string theory models. Specifically, one 
can have $M_c\simeq 2.6\times 10^{15}$ GeV, with new string-theoretic states 
affecting the 5d field theory regime at about $20M_c$. This is marginally 
consistent with the concept of a 5d orbifold GUT. 

Our conclusions are given in Sect.~\ref{conc}.

\section{The weak coupling regime of the heterotic string}\label{wc}
It is the purpose of this section to characterize the boundary of the 
perturbative regime of 10d heterotic string theory quantitatively. Although 
this is a familiar issue and the possible ways to achieve this are fairly 
obvious, we find it useful to include a detailed and hopefully 
self-contained discussion, where we carefully keep all relevant numerical 
coefficients. The results are crucial for the remainder of the paper since 
the potential loss of control of the stringy UV completion will be the main 
limiting factor in achieving realistic parameters in the low-energy field 
theory.

\subsection{Field theory lagrangian and string parameters}
The dynamics of heterotic string theory, at energies far below the mass of 
the first excited state, is described by the 10d supergravity plus super 
Yang-Mills lagrangian~\cite{Gross:1985rr,Polchinski:1998rr}
\beq
\label{actionH}
S_H=-\int d^{10}x\sqrt{G_H}e^{-2\phi_H} 
\left(\frac{1}{2 \kappa_{10}^2} R_H+\frac{1}{4\, g_{10}^2} {\rm Tr_V} 
F_{\mu\nu}F^{\mu\nu}\right)\,.
\eeq
Here the trace ${\rm Tr_V}$ is taken in the vector representation of SO(32) 
and the field strength is defined by $F_{\mu\nu}=-i\,[D_\mu,D_\nu]$ with $D_\mu 
=\partial_\mu+iA_\mu$. All fermionic and higher-dimension terms as well as 
the kinetic terms for the dilaton and 2-form field have been suppressed. 
Alternatively, the vector fields can also be defined to take values in the 
adjoint representation of SO(32), in which case one has to replace 
$\rm Tr_V$ with $(1/30){\rm Tr_A}$. This second form of the action then also 
describes the E$_8\times$E$_8$ heterotic theory. Note that the restriction 
of the above lagrangian to a regular SU($N$) subgroup of either SO(32) or 
E$_8\times$E$_8$ reads (with Minkowski metric and vanishing dilaton)
\beq
{\cal L}\supset -\frac{1}{2\, g_{10}^2} {\rm Tr_F} F_{\mu\nu}F^{\mu\nu}\,,
\eeq
where the trace is taken in the fundamental representation of SU($N$). Thus, 
for a compactification volume $V$, the conventional 4d GUT coupling (with 
phenomenological value $\sim 1/25$) is given by $\alpha_{\rm GUT}=g_{10}^2/ 
(4\pi V)$. 

The above action contains the two dimensionful parameters $\kappa_{10}$ and 
$g_{10}$, which are linked to the fundamental scale $\alpha'_H$ of the 
underlying string theory by~\cite{Gross:1985rr,Polchinski:1998rr
}~\footnote{
There appears to be a factor-of-2 discrepancy with~\cite{Witten:1996mz}.
}
\beq
\frac{\kappa_{10}^2}{g^2_{10}}=\frac{\al_H}{4}\,.\label{alh}
\eeq
Here we use standard conventions with a world sheet action $1/(4\pi 
\alpha'_H)\int d^2\sigma(\partial X)^2$, corresponding to a lowest-lying 
physical excited state with mass $m_H=2/\sqrt{\al_H}$. 

The dimensionless ratio $u\equiv g_{10}^4/\kappa_{10}^3$ is not a true 
parameter of the theory since it can be changed by shifting $\phi_H$ by a 
constant and then redefining $g_{10}$ and $\kappa_{10}$ so that the form of 
Eq.~(\ref{actionH}) is not modified. This exercise also demonstrates that
increasing the loop expansion parameter of string theory $g_H\sim 
\exp(\phi_H)$ by a given factor is equivalent to increasing $u$ by the 
same factor. Thus, for any given vacuum of the heterotic string, we are 
free to choose conventions where $\phi_H=0$ and to use $g_{10}$ and 
$\kappa_{10}$ or, equivalently, $\alpha'_H$ and $u$ as the unambiguous 
characteristics of this vacuum. The parameter $u$ then defines the strength 
of the string coupling. 

Another perspective on the situation, which may be particularly natural for 
a low-energy field theorist, is the following. The gravitational and gauge 
parts of the action of Eq.~(\ref{actionH}) (with $\phi_H=0$) contain the two 
mass scales $\bar{M}_{P,10}$ and $M_{YM,10}$, defined by 
\beq
\bar M_{P,10}^8=\frac{1}{\kappa_{10}^2}\,,\quad M_{YM,10}^6=\frac{1}{g_{10}^2
},
\eeq
which characterize the strong coupling regimes of the two respective field 
theories. On the one hand, a certain dimensionful combination of these two 
masses fixes, according to Eq.~(\ref{alh}), the string scale. On the other 
hand, the dimensionless combination
\beq
u=\left(\frac{\bar M_{P,10}}{M_{YM,10}}\right)^{12}
\eeq
determines the loop expansion parameter of the underlying string theory. 
Note that small heterotic string coupling corresponds to small 10d gauge 
coupling (measured in units of the 10d Planck mass). 

Note that the numerical coefficient in the proportionality relation $u\sim 
g_H$ derived above is, at the moment, arbitrary (as is the coefficient in 
the relation $g_H\sim\exp(\phi)$.) In the following, we will characterize 
the perturbative regime directly in terms of $u$ and, after this is done, 
we will give a definition of $g_H$ in terms of $u$ such that, to the best 
of our knowledge, perturbation theory breaks down at the special point 
$g_H=1$. (It is this specific definition of a string coupling $g=g_H$ that 
appears in Eq.~(\ref{fe}).) 

In principle, it should be possible to define a more fundamental string 
loop expansion parameter in terms of the field (which differs from $\phi$ by 
an unknown constant) multiplying the Euler term in the world sheet action. 
This parameter would then directly measure the world sheet topology.\footnote{
Note,
however, that this would require a consistent calculation of, e.g., the 
sphere and torus amplitude with a mutually compatible UV regularization.
}
In praxis, one starts in string perturbation theory by connecting the 
tree-level amplitude to field theory parameters. The one-loop amplitude 
is then related to the tree-level amplitude by a unitarity-based sewing 
procedure (see, e.g.~\cite{DiVecchia:1996uq}), and the relation to the 
Euler term of the world sheet action never arises. Comparing one-loop and 
tree-level terms one can then try to determine the field theory parameters 
characterizing the boundary of the perturbative regime and translate them 
into some (arbitrarily normalized) $g_H$. Conceptually, this is not very 
different from our approach based on the field-theoretically defined $u$. 
Of course, if one was given a number of non-trivial terms in the 
string-theoretic loop expansion of some physical quantity, a more 
trustworthy definition of perturbativity than the one we will offer could
be obtained. 

We finally note the useful expression 
\beq
m_H=\bar M_{P,10}\,u^{1/4}\label{mh}
\eeq
for the mass of the lowest-lying massive string state.

\subsection{Duality criterion for weak coupling of SO(32)}
One possibility to quantify the boundary of the weak coupling regime of the 
SO(32) heterotic string is to use the strong/weak coupling duality with type 
I string theory~\cite{Polchinski:1995df}. In analogy to Eq.~(\ref{actionH}), 
the relevant part of the low-energy action of the type I string reads
\bea
\label{actionI}
S_I=-\int d^{10}x\sqrt{G_I}
\left(\frac{1}{2 \kappa_{10}^2}e^{-2\phi_I}R_I+\frac{1}{4 g_{10}^2}
e^{-\phi_I}{\rm Tr_V} F_{\mu\nu}F^{\mu\nu}\right).
\eea
With the identifications $(G_I)_{\mu\nu}=e^{-\phi_H}(G_H)_{\mu\nu}$ and 
$\phi_I=-\phi_H$, 
Eqs.~(\ref{actionH}) and (\ref{actionI}) describe the same field theory. 
Duality means that, for any given set of parameters of this Yang-Mills 
supergravity action, the two possible UV completions via heterotic or type 
I string theory are, in fact, identical. 

The Regge slope $\alpha'_I$ of the type I string relates to the world sheet 
action as in the heterotic case above. This implies that the first excited 
state has mass $m_I=1/\sqrt{\al_I}$ and the field theory parameters are 
related to $\alpha'_I$ by~\cite{Sakai:1987jg,Tseytlin:1995fy,
Polchinski:1998rr}\footnote{
There 
appears to be a factor-of-2 discrepancy with~\cite{Caceres:1996is}.
}
\bea
\frac{g^2_{10}}{\kappa_{10}}=2\,(2\pi)^{\frac{7}{2}} \al_I.
\eea
We will describe the vacuum by $\phi_H=\phi_I=0$ such that 
Eqs.~(\ref{actionH}) and (\ref{actionI}) take the same form and continue 
to use the parameters $M_{P,10}$, $M_{YM,10}$ as well as the parameter 
$u$ characterizing the strength of the string coupling: $g_I\sim 1/u$. 
The mass of the lowest-lying excited string state is then given, in 
terms of the field theory parameters, by
\bea
m_I=\bar{M}_{P,10}\,\sqrt{2}(2\pi)^{7/4}\,u^{-1/4}\,.\label{mi}
\eea

Now, comparing Eqs.~(\ref{mh}) and (\ref{mi}), the following quantitative 
description of the weak and strong coupling regimes of the two string 
theories suggests itself. Taking, for example, $\bar{M}_{P,10}$ and $u$ as 
the fundamental parameters, focus first on the region where $u$ is very 
small and we can trust the perturbative heterotic description. According to 
Eqs.~(\ref{mh}) and (\ref{mi}), in this regime $m_H\ll m_I$. This implies 
that the perturbative type I description has broken down since there exists 
a massive state much lighter than the first excited state of string 
perturbation theory. Next, focussing on the region of large $u$, where 
$m_I\ll m_H$ and we know that we can trust type I perturbation theory, the 
same argument can be made for the heterotic string: The heterotic 
perturbative description has broken down since the type I excited state 
with mass $m_I$ (which must be non-perturbative in nature from the heterotic 
perspective) is much lighter than the lowest-lying massive heterotic state. 
Finally, if one attempts to extend the heterotic and type I perturbative 
descriptions to larger/smaller $u$ respectively, one obviously faces a 
fundamental problem at the point where $m_H=m_I$ since, at this point, 
perturbative and non-perturbative states with comparable mass exist from 
both perspectives. It is then natural to define this special (critical) 
point
\beq
u_c^2=4(2\pi)^7\label{uc}
\eeq
as the upper/lower boundary of the heterotic/type I perturbative 
regime.\footnote{ 
In ref.~\cite{Caceres:1996is}, a heterotic string coupling $\lambda_H$ is
defined and the actual expansion parameter of string theory is stated to be 
$\lambda_H^2/(2\pi)^5$. Setting this parameter to one corresponds, in our 
conventions, to a critical parameter $u^2_{c, {\rm Caceres\,\, et\,\,al.}}= 
(2/\pi^2) u_c^2.$ We can not comment on the origin of this small discrepancy 
since no derivation of the string expansion parameter appears 
in~\cite{Caceres:1996is}.
}

\begin{figure}[!t]
\hspace{2cm}
\includegraphics[scale=0.6]{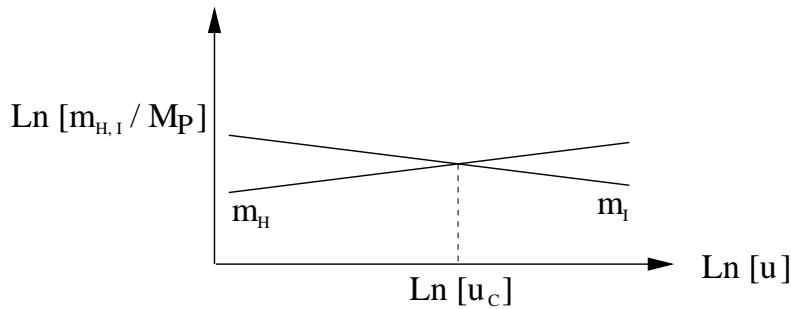}\vspace{-.3cm}
\caption{\small \it This figure shows $m_H$ and $m_I$ as functions of the 
coupling $u$ as defined by Eqs.~(\ref{mh}) and (\ref{mi}). The critical 
point $u_c$ separates heterotic (left) and type I (right) weak coupling 
regimes.}
\label{dual1}
\end{figure}

This situation is illustrated in Fig.~\ref{dual1} where the masses of 
the lowest-lying heterotic and type I state are plotted as functions of $u$. 
Note that, on the basis of perturbation theory, one can only trust the 
lower of the two lines at any given point $u$. In particular, the extension 
of the $m_I$ curve into the region of small $u$ is, as far as we know, 
only speculative at the quantitative level. However, the extension of the 
$m_H$ curve into the region of  large $u$ is known to be meaningful and to 
represent the heavy $D1$-brane of the type I theory~\cite{Polchinski:1995df}. 

As a cautionary remark, our characterization of the heterotic weak coupling 
regime by $u\ll u_c$ is certainly optimistic: We know that perturbation 
theory breaks down at $u=u_c$, but we can not exclude that it breaks down 
earlier. Note that so far we have only been concerned with perturbative 
calculability of the massive spectrum - the field theoretic gauge couplings 
will be discussed below.

\subsection{Duality criterion for weak coupling of E$_8\times$E$_8$}
A similar analysis can be performed for the E$_8\times$E$_8$ heterotic 
theory. The low-energy lagrangian of this theory is the same as in the 
SO(32) case, but the dual theory is now heterotic M-theory~\cite{
Horava:1995qa}, i.e., an $S^1/Z_2$ orbifold reduction of 11d supergravity 
with gauge fields localized at the fixed points. The parameters $\kappa_{10}$ 
and $g_{10}$ of the resulting 10d Yang-Mills supergravity lagrangian 
(cf.~Eq.~(\ref{actionH}) with $\phi_H=0$) are related by 
\beq
\label{Mth-E8}
\kappa_{10}^2=\frac{\kappa_{11}^2}{2\pi \rho}\qquad\mbox{and}\qquad
g_{10}^2=2\pi(4\pi\kappa_{11}^2)^{2/3}
\eeq
to the 11d gravitational coupling $\kappa_{11}$ and the radius $\rho$ of the 
$S^1$. Note that, following~\cite{Horava:1995qa}, these relations assume 
compactification on an $S^1$ with $Z_2$ symmetry of the bulk fields (rather 
than compactification on an interval of size $\pi\rho$ with appropriate 
boundary conditions). 

\begin{figure}[!t]
\hspace{1.4cm}
\includegraphics[scale=0.6]{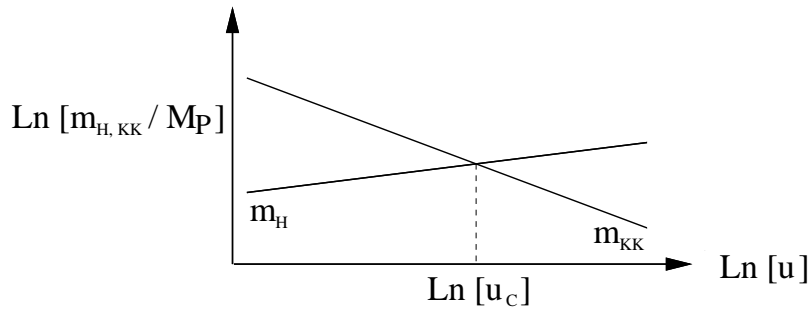}\vspace{-.3cm}
\caption{\small \it This figure shows $m_H$ and $m_{KK}$ as functions of the 
coupling $u$ as defined by Eqs.~(\ref{mh}) and (\ref{mthu}). The critical 
point $u_c$ separates heterotic-string (left) and 11d-supergravity (right) 
weak coupling regimes.}\label{dual2}
\end{figure}

Although, in contrast to the type I string, this dual theory does not 
contain a microscopic description (i.e., an explicit UV completion), we can 
nevertheless repeat the strong/weak coupling argument of the last subsection. 
Indeed, since $\rho\sim u^{3/4}/\bar{M}_{P,10}$, we can trust the 11d field 
theory description at large $u$. This implies that the lowest-lying massive 
state in this regime is the first Kaluza-Klein (KK) excitation of 
supergravity with mass $m_{KK}=1/\rho$. In our definition of perturbativity 
this mass will take the place of $m_I$ in the last subsection. Thus, as 
before, we will assume that the E$_8\times$E$_8$ heterotic string is weakly 
coupled as long as $m_H$ (the same as in the SO(32) theory) is smaller 
than $m_{KK}$. Explicitly, we have 
\bea
\label{mthu}
m_{KK}=\bar{M}_{P,10}\,2(2\pi)^{7/2}\,u^{-3/4}\,,
\eea
which is illustrated in Fig.~\ref{dual2}. Although Eq.~(\ref{mthu}) is very 
different from the analogous relation, Eq.~(\ref{mi}), the critical point 
$m_H=m_{KK}$ lies, intriguingly, at exactly the same value of $u$, 
$u_c^2=4 (2\pi)^7$, as in the SO(32) case\footnote{Note that a different 
relation between 11d and 10d parameters was found in~\cite{Conrad:1997ww}, 
corresponding to an extra factor $2^{1/3}$ in the expression for $g_{10}^2$ 
in Eq.~(\ref{Mth-E8}) and implying the somewhat less restrictive value 
$u_c^2=8 (2\pi)^7$.}. 

This motivates us to give a precise definition of a heterotic and a type I 
string coupling, $g_H$ and $g_I$,
\beq 
g_H=\frac{1}{g_I}=\frac{u}{u_c}=\frac{1}{2(2\pi)^{7/2}}\left(\frac{ 
\bar{M}_{P,10}}{M_{YM,10}}\right)^{12}\,.\label{scd}
\eeq
These couplings relate in the usual way to the dilaton field, $g_H\sim\exp(
\phi_H)$, have a clear duality based definition in that $g_H=g_I=1$ 
characterizes the crossing points in Figs.~\ref{dual1} and \ref{dual2}, and 
will simplify subsequent formulae involving the 4d gauge coupling 
$\alpha_{GUT}$. Most importantly, with these definitions the usual 
perturbativity requirements $g_H<1$ and $g_I<1$ now have a more precise 
motivation in terms of the correctly predicted lowest-lying massive states.

\subsection{Naive dimensional analysis in effective field theory}
It is also possible to estimate the critical value of $u$, where 
string-theoretic perturbativity is lost, directly within the 10d low-energy 
effective field theory. More specifically, we will approximate the 
string-theoretic one-loop effect by the field-theoretic one-loop integral cut 
off at the string scale. Equivalently, we can take the point of view that 
the 10d gauge theory with cutoff $\Lambda$ has a dimensionless expansion 
parameter $\sim g_{10}^2\Lambda^6$ and require the cutoff (i.e. the string 
scale) to be sufficiently low to ensure that this expansion parameter is 
smaller than one. 

To be generic, we begin with the Yang-Mills lagrangian in $d$ dimensions 
written in a representation $R$,
\beq
{\cal L}_{YM}=-\frac{1}{4 g^2_{d,R}} {\rm Tr_R} 
F_{\mu\nu}F^{\mu\nu}\,,\label{ymlag}
\eeq
where the field strength is defined as after Eq.~(\ref{actionH}). On the 
basis of this lagrangian, the ratio of one-loop and tree-level terms 
(e.g., the one-loop correction to the gauge boson propagator) can be 
estimated as
\beq
\frac{\rm one\,\,loop}{\rm tree\,\,level}\,\simeq\,g_{d,R}^2\left(\frac{ 
\Lambda^{d-4}}{(d\!-\!4)2^d\pi^{d/2}\Gamma(d/2)}\right)\,\left(\frac{C_A\, 
d(A)}{C_R\,d(R)}\right)\,.\label{10}
\eeq
Here the first factor $g_{d,R}^2$ is obvious, the second factor comes from 
the standard loop integral~\cite{Chacko:1999hg}
\beq
\int^\Lambda\frac{d^dk}{(2\pi)^d}\,\frac{1}{(k^2)^2}\simeq\frac{1}
{2^d\pi^{d/2}\Gamma(d/2)}\int_0^\Lambda|k|^{d-5}d|k|\simeq\frac{\Lambda^{d-4}
}{(d\!-\!4)2^d\pi^{d/2}\Gamma(d/2)}\,,
\eeq
while the last factor arises from the group-theoretic degeneracy of the 
gauge bosons propagating in the loop. It can be understood as follows
(see, e.g., ref.~\cite{vanRitbergen:1998pn} for group-theoretic details):
The inverse propagator derived directly from Eq.~(\ref{ymlag}) (without 
canonical field redefinition) comes with a factor $T_R$, defined by Tr$_R({ 
\bf T}_a{\bf T}_b)=T_R \delta_{ab}$ for group generators ${\bf T}_a, {\bf 
T}_b$. The 3-gauge-boson vertex comes with a factor $T_Rf_{abc}$, where 
the structure constants are defined by $[{\bf T}_a,{\bf T}_b]=if_{abc}{\bf 
T}_c$. Thus, the one-loop correction to a propagator, which contains two 
extra vertices and three extra propagators, has a relative group theory 
factor $C_A/T_R$, where $C_A$ is defined by $f_{abc}f_{dbc}=C_A\delta_{ad}$. 
(More generally, $C_R$ is the quadratic Casimir of the representation $R$, 
$({\bf T}_a)_{ij}({\bf T}_b)_{jk}=C_R\delta_{ik}$.) Using the identity
$T_R\,d(A)=C_R\,d(R)$, where $d(A)$ and $d(R)$ are the dimensions of the 
adjoint representation and the representation $R$ respectively, one then 
finds the last factor in Eq.~(\ref{10}). 

We would like to make some technical comments concerning this 
group-theoretical loop-factor: Clearly, we can not expect this ``colour 
factor'' to be exact -- it is only meant to capture the dominant effect for 
large groups (e.g., the large $N$ limit of SU($N$)) and can be changed by 
an ${\cal O}(1)$ multiplicative constant if different conventions are chosen. 
In our conventions, the colour factor becomes trivial in the adjoint 
representation, which we consider a natural choice. Also, if we consider 
4d SU($N$) with the lagrangian of Eq.~(\ref{ymlag}) and $R=F$ the fundamental 
representation, the expansion parameter of Eq.~(\ref{10}) becomes
$2g_{4,F}^2N/(16\pi^2)$. In this case, the conventional gauge coupling used in 
phenomenology is defined as $g^2=2g_{4,F}^2$. Thus, the loop expansion 
parameter is equal to the 't Hooft 
coupling $g^2N$ divided by the familiar 4d loop factor $16\pi^2$. For 
$N\!=\!3$, the expansion parameter is then $3\alpha_s/(4\pi)$, which is 
consistent with what is found in explicit QCD loop calculations. 

Returning to the SO($N$) case relevant for the heterotic string, the colour 
factor appearing in Eq.~(\ref{10}) is $N\!-\!2$, i.e. 30 for SO(32). This 
is also the correct factor 
for E$_8\times$E$_8$, as is obvious from the normalization factor (1/30) 
mentioned after Eq.~(\ref{actionH}) (which we now have effectively derived). 
Thus, with $d=10$ and $\Lambda=m_H$, the requirement that the loop factor 
of Eq.~(\ref{10}) is equal to one leads to a condition on $u$. The 
resulting critical value
\beq
u_c'^2=\frac{3}{5}\cdot 2^8\cdot(2\pi)^5
\eeq
is astonishingly close to that of Eq.~(\ref{uc}), $u_c'^2/u_c^2=48/(5\pi^2) 
\simeq 0.97$. This should presumably not be taken too seriously, given that 
$\Lambda=m_H$ was an ad-hoc choice and that a small variation of the cutoff 
would affect the result significantly. Nevertheless, we view this subsection 
as an important confirmation that the critical coupling $u_c$ obtained before 
has the correct order of magnitude and that, with the above definition of the 
string coupling, $g_H<1$ is a useful criterion for perturbativity. We hope 
that this quantitative definition of the perturbative regime will also prove 
useful in other settings (cf. e.g. the discussion 
in~\cite{Antoniadis:2004dt}).

\section{Isotropic vs. anisotropic compactifications}\label{ia}
\subsection{The isotropic case}\label{iso}
As discussed in detail in~\cite{Caceres:1996is}, isotropic compactifications 
(with all the compactification radii roughly equal) do not straightforwardly 
lead to realistic unified models, neither in heterotic nor in type~I string 
theory (with only $D9$-branes present). A crucial assumption made in this 
context is that the unification scale is linked to either the string or 
the compactification scale. The problems are due to the phenomenological 
requirement $\alpha_{GUT}\simeq 1/25$ together with the compactification 
formulas $\alpha_{GUT}=g_{10}^2/(4\pi V)$ and $M_P^2=8\pi \bar{M}_P^2=8\pi 
\bar{M}_{P,10}^8V$, with $M_P=1.22\times 10^{19}$ GeV. (We base our analysis 
on 1-loop MSSM running above an effective SUSY breaking scale $\sim m_Z$, 
leading to gauge coupling unification at $M_{GUT}\simeq 2\times 10^{16}$ 
GeV.)

To be specific, in a toroidal system with $V=(2\pi R)^6$, the gauge 
coupling is given by 
\beq
\frac{\alpha_{GUT}}{2}=\frac{g_H^2}{(Rm_H)^6}=\frac{g_I}{(Rm_I)^6}\,,
\label{algut}
\eeq
with the string couplings defined in Eq.~(\ref{scd}). Although written down
in the oversimplified case of a torus, these equations can also be used 
to discuss the more general Calabi-Yau or orbifold case with typical length
scale $R$. The above relations are to be supplemented with the formulae
\beq
M_P^2\frac{\alpha_{GUT}}{2}=m_H^2=g_I^{-1}m_I^2\,,\label{mp}
\eeq
fixing the string scale in terms of the Planck scale and leading famously 
to the $g_H$-independent result $\alpha'^{-1/2}_H=8.6\times 10^{17}$ GeV 
in the heterotic case. 

The unification scale problem is now very transparent. Focus first on the 
heterotic case and take the (arguably natural) radius $R\simeq m_H^{-1}$. The 
parametric smallness of $\alpha_{GUT}$ is then equivalent to a small 
string coupling $g_H$, which justifies quantitative gauge coupling 
unification, but at the too high scale $m_H$. In the type I case, we
can analogously assume $R\simeq m_I^{-1}$, implying small $g_I$ and leading
to the somewhat better but nevertheless high unification scale $m_I\simeq 
M_P\alpha_{GUT}/2\simeq 2.4\times 10^{17}$ GeV. Of course, we are using 
only the simplest form of heterotic-type I duality, considering type I
theory with just $D9$-branes. This is a very restricted framework where the 
whole wealth of simple orbifold models of the heterotic theory is not 
available. One would then have to stick to the true Calabi-Yau case, which is 
difficult since the radius is too small to justify the supergravity 
approximation. Considering type I theory on its own, many interesting 
$D$-brane models can and have been considered. However, this goes beyond 
the scope of this work, which focusses on geometric constructions based 
directly on the 10d super Yang-Mills theory and its stringy UV completion. 

The alternative point of view is to take $g_H=g_I=1$ and to derive the 
parametric smallness of $\alpha_{GUT}$ from a large compactification 
volume. This corresponds to identifying $M_{GUT}$ with $R^{-1}$. However, 
due to the high power of $R$ appearing in Eq.~(\ref{algut}), the improvement 
is only minor: $R\simeq 2m_{H,I}$ is sufficient to explain the observed 
small gauge coupling so that the unification scale, if identified with 
$R^{-1}$, is not lowered significantly. 

We would now like to comment on the heterotic M-theory case dual to the 
E$_8\times$E$_8$ heterotic string. To discuss this case, Eqs.~(\ref{algut})
and (\ref{mp}) have to be supplemented with the relation $\alpha_{GUT}/2= 
g_H^{-4}/(Rm_{KK})^6$ and $M_P^2\alpha_{GUT}/2=g_H^2m_{KK}^2$. In analogy 
to the heterotic and type I string theory discussion above, we can explain 
the smallness of $\alpha_{GUT}$ by either small $g_H^{-4}$ or large $R$. 
However, neither option leads to a sufficiently small GUT scale (if this 
scale is to be identified with $R^{-1}$). The interesting new possibility 
arising in the M-theory case is to take $R\ll m_{KK}^{-1}$ and $g_H^{-4}\ll 
1$, which does indeed allow for solutions with $R^{-1}\ll M_P$. Nevertheless, 
limitations arise since the product space structure of 6d manifold and 
$S^1/Z_2$ assumed above is lost~\cite{Witten:1996mz}.

\subsection{Taking some of the radii large}\label{lr}
The possible solution to the above difficulties on which we now want to 
focus uses compactifications with very different radii such that the GUT 
scale is linked to the larger radii. As will be discussed below in 
explicit examples, this can be realized if the GUT group is broken by 
Wilson lines whose length is associated with those radii. 

To be specific, consider heterotic models with internal space $T^d\times 
T^{6-d}$, where $T^d$ is symmetric with large radii $R_l$, and $T^{6-d}$ is 
symmetric with small radii $R_s$. Equation~(\ref{algut}) is now replaced by
\beq
\frac{\alpha_{GUT}}{2}=\frac{g_H^2}{(R_lm_H)^d(R_sm_H)^{6-d}}\,.\label{alg}
\eeq
As before, we use this simple toric situation as an approximation to the 
more interesting asymmetric Calabi-Yau or orbifold case. 

The most favourable case of $d=1$ leads, for $g_H\simeq 1$ and $R_s\simeq 
m_H^{-1}$, to a GUT breaking scale $R_l^{-1}\simeq m_H/50\simeq 
3.4\times 10^{16}$ GeV. In this situation, one is effectively considering 
a 5d theory in some intermediate energy range. To have a non-contractible 
Wilson line, this must be compactified to 4d on an $S^1$. Unfortunately, 
this situation can not be realized with Calabi-Yaus, which do not possess
1-cycles. It can also not be realized in the toric orbifold case since the 
$S^1$ compactification implies unbroken ${\cal N}\!=\!2$ SUSY 
in 4d.\footnote{
This 
is easy to see since one is effectively considering a 5d theory 
compactified on $S^1$. If the 5d SUSY corresponds to ${\cal N}\!=\!2$ (in 
4d language), the allowed SU(2) R-symmetry can only break this SUSY to 
${\cal N}\!=\!0$ or not at all. If the 5d SUSY is ${\cal N}\!=\!4$, the 
small transverse space must be free of singularities. Then a breaking 
to ${\cal N}\!=\!1$ in 4d would mean that one has explicitly found 
a locally flat true Calabi-Yau, which is known not to exist.
}
Thus, the lack of promising examples forces us to drop the $d=1$ case in this 
context. 

The case $d=1$ remains interesting in the following sense: Even if $R_l$ is 
not the GUT scale, the relative logarithmic running of inverse SM gauge 
couplings could be different above $R^{-1}$ allowing the true GUT scale 
(where this differential running stops) to be changed. This 
occurs naturally in the context of 5d or 6d orbifold GUTs~\cite{Hall:2001pg, 
Hebecker:2001wq}. Large threshold corrections producing such an effect where discussed
early on in string model building~\cite{Nilles:1997vk}.
Of course, such scenarios with modified differential 
running above the scale $R_l^{-1}$ can also be consider for $d>1$. 

In the second best case of $d=2$, we will show explicitly below that, even 
at the simplest level of toric orbifolds, examples with 4d ${\cal N}\!=\!1$ 
SUSY and GUT-breaking Wilson lines associated with the two large radii exist. 
Numerically, the situation is clearly worse than for $d=1$, since $g_H\simeq 
1$ and $R_s\simeq m_H^{-1}$ implies $R_l^{-1}\simeq m_H/7\simeq 2.5\times 
10^{17}$ GeV. Alternatively, we can insist on $R_l^{-1}\simeq M_{GUT}$, 
implying the constraint $g_H/(R_sm_H)^2\simeq 12$, which means that we 
have to face either small radii below the string length scale or a large 
string coupling.

\subsection{Facing very small radii or a large string coupling}\label{fac}
As argued above, we are led to consider string compactifications to 
$4+d$ dimensions where $R_s$ and $g_H$ are constrained by Eq.~(\ref{alg}). 
We emphasize that, in this equation, $\alpha_{GUT}$, $m_H$ and $R_l$ are 
now fixed by phenomenology ($R_l^{-1}\simeq M_{GUT}$ in the simplest 
case but possibly varying depending on the details of the 
running). The interesting question is then how much quantitative control 
over this $4+d$ dimensional model we can gain and how much control we need in 
order to ensure quantitative gauge coupling unification. As already explained 
in the most promising case of $d=2$, the fully perturbative regime $g_H\ll 1$ 
and $R_s\gg m_H^{-1}$ is excluded. 

The crucial point to be made in this subsection, which is also at the heart 
of the models with non-local GUT breaking discussed later on, is that 
we do not need a perturbative UV completion of the $4+d$ dimensional 
field theory. Specifically, in the case of $d=2$ and non-local GUT breaking, 
we only require control over light 6d states up to the scale $R_l^{-1}$. The 
contribution of all heavier states to non-universal GUT threshold 
corrections will be exponentially suppressed because of field theoretic 
locality. This is based on the fact that, on length-scales smaller than 
$R_l$ in the 6d field theory, the GUT group is never broken. 

However, the absence of such dangerous light states is not entirely obvious. 
In particular, winding modes with mass $\sim R_s/\alpha'$ are in general 
present. They become light in the limit of small $R_s$. This in itself would 
not be a very serious difficulty since, for $d=2$ and $g_H={\cal O}(1)$, 
small radii $R_s\simeq m_H^{-1}/\sqrt{12}$ are sufficient. Thus, the winding 
states are much heavier than $M_{GUT}$. The more serious issue is that these 
light states can spoil string-theoretic perturbativity. This can be 
quantified most easily by appealing to $T$ duality. In the simplest case, 
$T$ duality means that a physical setting with Regge slope $\alpha'_H$, 
string coupling $g_H$, and a single compactification radius $R$ can be 
equivalently described by a string theory with parameters $\alpha'_H$, 
$g'_H=g_H\sqrt{\alpha'}/R$ and compactification radius $R'=\alpha'/R$~\cite{ 
Buscher:1987qj}. The above light winding states are the KK states of the dual 
description. The real problem is the potentially very large dual string 
coupling that arises if the duality is applied to all small radii. This 
strongly coupled theory may contain extremely light non-perturbative 
states. 

Before entering a more detailed discussion of specific settings, we have to
recall that we used and will continue to use tree level formulae in our 
coupling strength analysis. This is justified if the intermediate-energy 
$4\!+\!d$ dimensional theory has maximal (${\cal N}=4)$ SUSY protecting 
us from loop corrections and if we can appeal to bulk dominance (parametrical 
largeness of the $4\!+\!d$ dimensional volume) in the last compactification 
step. If, at the intermediate stage, we encounter a higher-dimensional 
${\cal N}=2$ gauge theory, ${\cal O}(1)$ corrections to the gauge coupling 
may arise. We ignore them with the argument that, at the present level of 
precision, all that matters is the approximate size of the underlying string 
coupling and not the specific ${\cal O}(1)$ factor linking it to the GUT 
gauge coupling. However, it is conceivable that in very specific settings, 
e.g. in situations close to the UV fixed point regime of~\cite{
Seiberg:1996bd}, the 1-loop corrections of the ${\cal N}=2$ SUSY gauge theory 
modify the naive tree-level analysis dramatically and allow for a solution 
of the string scale/GUT scale problem very different from the one discussed 
in this paper.

\subsubsection{Two large and four small radii}\label{twolarge}
The potential problem of dangerous light states can be brought under control 
if, using a network of dualities~\cite{Antoniadis:1999rm}, we manage to find 
a perturbative UV completion in the simplified case 
of a flat $4\!+\!d$ dimensional theory. This may be possible 
by going to the type I description via $S$ duality. We now focus on $d=2$. 
Consider the two-dimensional space parameterized by $g_H$ and $R_s$ and 
enforce the constraint 
\beq
g_H=12(R_s m_H)^2\,.\label{con}
\eeq
This defines a one-parameter set of models within which we can search for an 
appropriate dual description of our 6d low-energy theory. The type I duals 
have parameters $g_I=g_H^{-1}$ and $m_I=g_H^{-1/2}m_H$. Unfortunately, we 
immediately find $R_sm_I\simeq 1/\sqrt{12}$, i.e., we have the problem 
of too small radii. Notice that the product of dual string scale and small 
radius does not depend on the choice of parameters in the original heterotic 
model. This is a special feature of the case $d=2$. Next, we $T$-dualize the 
type I model. The resulting $T$-dual type IIB theory has coupling $g_{II}= 
g_I\alpha_I'^2/R_s^4$, radii $R_s'=\alpha_I'/R_s$ and gauge fields localized 
on $D5$-branes~\cite{Dai:1989ua}. Writing this in terms of the original 
parameters and expressing $R_s$ as a function of $g_H$ and $m_H$, one finds
\beq
g_{II}\simeq (12)^2g_H^{-1}\qquad\mbox{and}\qquad R_s'\simeq\frac{\sqrt{12 
g_H}}{m_H}\,.\label{dr}
\eeq
Using $g_H\simeq (12)^2$ to keep $g_{II}$ at the boundary of perturbativity, 
this implies $R_l/R_s'\simeq 2$, i.e., the KK states of this (almost)
perturbatively controlled model are marginally above the GUT-breaking scale. 
Notice that all other string states (e.g., the fundamental string excitations 
with mass $m_I$) are heavier than these KK states and are therefore less 
problematic.

Of course, we do not really need the type II model with coupling $g_{II}$ to 
be perturbative. Our interest is only in the light states of mass $R_s'^{-1}$.
We can therefore go to smaller $g_H$, hoping that Eq.~(\ref{dr}) remains 
qualitatively valid and the dangerous light states become heavier. The most 
we can do is to take $g_H\simeq 48$ since, going to even smaller $g_H$, the 
coupling $g_H'$ of the heterotic $T$ dual becomes larger, $g_H'>g_H$ 
(using the constraint Eq.~(\ref{con})). Equation~(\ref{dr}), if taken 
seriously, then implies $R_l/R_s'\simeq 3.5$.

The presence of the light (a factor of 2 to 3 above $R_l^{-1}$) 
non-perturbative states found above is potentially dangerous. However, since 
they are Kaluza-Klein excitations in the internal dimensions orthogonal
to the $D5$-branes, they are not charged under the gauge group. We can 
therefore hope that their sole effect will be a mild modification of the 
internal geometry, not affecting the basic picture of field-theoretic 
non-local GUT breaking at $M_{GUT}\simeq R_l^{-1}$. Furthermore, since 
field-theoretic locality leads to an {\it exponential} suppression of all
effects related to states above the scale $R_l^{-1}$, we can hope that 
the very mild hierarchy $R_l/R_s'$ found above is sufficient. 

In many concrete models, the situation will be further improved by the 
following effect. In the presence of Wilson lines, which are anyway a crucial 
ingredient in realistic model building, the toroidal dimensional 
reduction~\cite{Narain:1986am} and, consequently, the T-duality rules~\cite{ 
Hassan:1995pk} are modified, with the effect that the potential 
perturbativity loss can be avoided. 

Specifically, in a model with $g_H<1$ and $R_s<\al_H^{1/2}$, the potential 
loss of 
calculability is due to the light winding modes of mass $m\sim R_s/\al_H$. The 
presence of a Wilson line $A$ modifies the relevant mass formula. Focussing
for simplicity on the case of one compact dimension, the masses of states 
in a gauge representation labelled by the root-space vector $q$ 
read~\cite{Narain:1986am,Polchinski:1998rr} 
\beq
m^2=\left[\frac{n}{R}-q\cdot A-wR\left(\frac{1}{\al_H}+\frac{A\cdot A}{2}
\right)\right]^2+\frac{4\tilde{N}}{\alpha'_H}\,.
\eeq
Here the dot-product refers to vector multiplication in root space and the 
integers $n$ and $w$ are the KK and winding numbers, which are subject to 
the constraint
\beq
nw=1-N+\tilde{N}-\frac{1}{2}\,q\cdot q\,.
\eeq
The non-negative integers $N$ and $\tilde{N}$ characterize the excitation 
levels of the left-moving world-sheet bosons and the right-moving Ramond
sector respectively.

The crucial point following from the above equations is that, for a moderately 
large Wilson line value, the tower of light winding modes with 
spacing $R/\alpha'_H$ disappears.\footnote{
As 
an example, one can consider the case $A\sim\frac{1}{R}\times \frac{\sqrt{R}} 
{\al^{1/4}}$ and check that even for very small radii the lightest massive 
state has mass $\sim \alpha'^{-1/2}$.
}
At least naively, this seems to solve the problem completely. From a duality 
point of view this is also apparent since the modification of the $T$-duality 
rules~\cite{Hassan:1995pk} is such that the $T$-dual theory has a smaller 
coupling than without Wilson lines. As a result, we can expect that 
potentially dangerous non-perturbative states, whose mass grows with the 
inverse coupling, become heavier.

\subsubsection{One large and five small radii}
As already explained, the analysis is very sensitive to the number of large
compact dimensions. For $d>2$ the situation is qualitatively similar to the 
$d=2$ case but, clearly, the potential perturbativity loss is more severe. We
do not enter into a numerical discussion of these cases. The case $d=1$, 
instead, deserves more care. Indeed, even though we know that a single large 
internal dimension cannot be used to set the scale of non-local GUT breaking, 
it can provide low-scale thresholds which can be relevant for the 
GUT scale/string scale problem. 

Therefore, in the following discussion of the $d=1$ case, we do not set 
$R_l^{-1}$ equal to $M_{GUT}$ but keep it as a free parameter. We are 
thus studying a space parameterized by $R_l$, $R_s$ and $g_H$ with the 
constraint 
\beq
\frac{\alpha_{GUT}}{2}=\frac{g_H^2}{(R_lm_H)(R_sm_H)^{5}}\,, 
\eeq
which provides us with a two-parameter set of models. We parameterize this 
set by $R_l$ and $R_s$. As long as $R_l<50/m_H$, we can choose $R_s\simeq 
m_H^{-1}$ and find ourselves in the perturbative heterotic regime. This 
puts the lowest-lying massive states of the 5d field theory at the scale 
$m_H$. If, however, $R_l>50/m_H$, the heterotic UV completion becomes 
non-perturbative. 

In this case, we can try the type I dual with $g_I=g_H^{-1}$ and $m_I= 
g_I^{1/2}m_H$. One finds 
\beq
(R_sm_I)\,g_I^{-1/10}\simeq \left(\frac{50}{R_l m_H}\right)^{1/5}\,,
\eeq
implying that this description is also non-perturbative (either $g_I^{-1}$
or $R_sm_I$ have to be smaller than one). The $T$ dual of this type I model, 
which is a type IIA model with $D4$-branes, has parameters
\beq
R_s^\prime=\frac{\alpha'_I}{R_s}\simeq \frac{R_l}{7}\frac{(R_s m_H)^{3/2}}
{(R_l m_H)^{1/2}}\qquad\mbox{and}\qquad g_{II}=\frac{g_I}{(m_I 
R_s)^5}=\frac{(R_l m_H/50)^{3/4}}{(R_sm_H)^{5/4}}\,,
\eeq
which finally provides a perturbative UV completion. More specifically, we 
can go to the boundary of the perturbative domain, $g_{II}\simeq 1$, by 
choosing $R_s$ such that 
\beq
(R_s m_H)^{5/4}\simeq(R_l m_H/50)^{3/4}\,. 
\eeq
This implies 
\beq
R_s^\prime\simeq \frac{R_l}{50}\left(\frac{R_lm_H}{50}\right)^{2/5}\simeq 
m_I^{-1}\left(\frac{R_lm_H}{50}\right)^{2/5}\,,\label{r50}
\eeq
so that, from a 5d field theory perspective, the lowest-lying massive states 
appear at a scale $50M_c$ in the best case. As will be discussed in more 
detail in Sect.~\ref{locbrk}, this validity range of the 5d field theory is 
sufficient for quantitative gauge coupling unification in the orbifold GUT 
framework. However, $R_l$ can not be made much larger than $50/m_H$ without 
restricting the above validity range. (The second part of Eq.~(\ref{r50}) 
confirms that there is no danger from too small radii in the type IIA model.)

\subsubsection{Large and small radii in the E$_8\times$E$_8$ theory}
\label{radiiE8}
In the case of the heterotic E$_8\times$E$_8$ theory, we have less control 
than in the SO(32) case discussed above. The reason is that, after using the 
analogue of $S$ duality to go to the $M$ theory (11d supergravity on 
$S^1/Z_2$) 
side of the model, we do not have the tool of $T$ duality at our disposal. It 
is then difficult to quantify the danger of making the small radii $R_s$ too 
small. We attempt to achieve this in the following way.

From the formulae of Sect.~(\ref{iso}) one derives the $S^1$ radius
\beq
\rho=\sqrt{R_l^d R_s^{6-d}}\,\frac{m_H^2}{7}\,,\label{rho}
\eeq
with $m_H=2/\sqrt{\al_H}=2\times8.6\times 10^{17}$ GeV being fixed 
phenomenologically. To proceed, we start from the `critical point' 
$g_H=1$, where we assume both the heterotic string and the $M$ theory
description to be qualitatively correct. At this point, we set the 
small radii to the minimal value suggested by the string theory description, 
$R_s\simeq m_H^{-1}$. We then find $\rho\simeq R_l/7\simeq m_H^{-1}$ for 
$d=2$ and $\rho\simeq R_l/50\simeq m_H^{-1}$ for $d=1$. 

Now we want to see whether we can make $R_l$ larger, which corresponds to 
moving into the $M$ theory regime. We assume that the smallness of $R_s$ is 
limited by physics local in the 11th dimension (i.e., we are not 
allowed to make $R_s$ too small in units of $g_{10}^{1/3}$ or $\kappa_{11}^{
2/7}$). Thus, when moving away from $g_H=1$ the small radii, if kept at their 
minimal value, scale as
\beq
R_s\sim g_{10}^{1/3}\sim (R_l^dR_s^{6-d})^{1/6}\,,
\eeq
where the second proportionality follows from the phenomenologically fixed 
4d gauge coupling. The above implies $R_s\sim R_l$ so that, on the basis 
of Eq.~(\ref{rho}), the parameter $\rho$ grows like $R_l^3$ if $R_l$ 
is increased above its critical value (i.e., its value at $g_H=1$ and 
minimal $R_s$). From the low-energy field theory perspective, this means
that the threshold set by $\rho$ (and thus the potentially non-perturbative 
physics related to the interplay of the curved branes and the bulk) move 
quickly into the infrared if we attempt to take $R_l$ to or below the GUT 
scale. 

Numerically, for $d=2$ the ratio of small and large radii is fixed to 
$R_s/R_l\simeq 1/7$. The relation 
\beq
\rho\simeq \frac{R_l}{7}\left(\frac{R_lm_H}{7}\right)^2,
\eeq
then implies that, when $R_l$ is increased, one encounters the equality 
$R_l=\rho$ already at $R_l^{-1}\simeq 9.3\times 10^{16}$ GeV. Thus, 
it is impossible to reach the GUT scale without leaving the 6d gauge theory 
regime (although the mismatch is not too severe). 

In the $d=1$ case, instead, the ratio of radii is fixed at $R_s/R_l\simeq 
1/50$ and the $M$ theory radius is given by 
\beq
\rho\simeq \frac{R_l}{50}\left(\frac{R_lm_H}{50}\right)^2\,.
\eeq
The large power of $R_l$ on the r.h. side makes the situation less 
favourable than in the SO(32) case, where an analogous relation is given by 
the first part of Eq.~(\ref{r50}). For example, $\rho=R_l$ implies 
$R_l^{-1}\simeq 4.8\times 10^{15}$ GeV, meaning that there is no 5d field 
theory regime below that scale.\footnote{
Unfortunately, 
the present quantitative analysis does not confirm the more optimistic 
conclusions of the related order-of-magnitude discussion 
in~\cite{Hall:2001xb}. 
}
Even with $R_l^{-1}\simeq M_{GUT}$ and the 
resulting cutoff $\rho^{-1}\simeq 3.5\times 10^{17}$ GeV the validity range 
of 5d field theory is still relatively small (a factor $\sim 17$ in energy 
scales). 

In both cases, the constraints on low values of $R_l^{-1}$ and a 
corresponding higher-dimensional field theory at low energy scales 
are more severe than for the SO(32) heterotic string. However, our 
analysis was conservative: larger $R_l$ might be feasible if, for example, 
$R_s$ could be taken to values below $m_H^{-1}$ (recall that this appears 
to be possible in the SO(32) case because of the modified $T$ duality in the 
presence of Wilson lines). We discuss in~\ref{OGUT5D} the relevance of these 
bounds on heterotic $E_8\times E_8$ orbifold GUT model building.

\section{Non-local vs. local gauge symmetry breaking in string theory}
\label{nln}
\subsection{Non-local breaking on a freely-acting orbifold}\label{nln1}
Orbifold compactifications of string theory allow for many gauge and SUSY 
breaking patterns which can be classified using the concept of  ``locality''. 
We will call the breaking of a certain symmetry in an extra-dimensional 
model ``local'' if it is realized at certain points in the internal space; 
it is instead ``non-local'' if it is due to some global property of the 
whole internal space. In the first case, the scale of breaking is tied to 
the cutoff, e.g., the string scale. In the second case, it is related to 
the size of the internal space. 

A standard example of local gauge symmetry and SUSY breaking is the
quantization on an orbifold generated by a discrete symmetry $g$ which 
acts as a rotation in the internal space, combined with a non-trivial 
transformation of the gauge bundle. The breaking is local since there are, 
in the internal space, points $x_f$ such that $g\,:\,x_f\to x_f$. The gauge 
symmetry and SUSY breaking is localized in these fixed points. Furthermore, 
the twisted states, which come in multiplets of the reduced gauge symmetry 
(SUSY) and do not fill out multiplets of the original symmetry, live at these 
points. 

Non-local breaking can instead be realized by modding out a discrete 
symmetry generated by an operator $h$ which acts freely (i.e., without fixed 
points) in the internal space. In this case the symmetry breaking due to 
$h$ is not localized and the breaking scale is not tied to the string scale.
Instead, it is set by some of the compact radii (cf. Sect.~\ref{ex1}).
For example, defining $h$ as a translation in the internal space, combined 
with a rotation in the gauge (R-symmetry) group, a non-local gauge symmetry 
(SUSY) breaking can be realized. This mechanism is also known as 
Scherk-Schwarz symmetry breaking~\cite{Scherk:1979zr} or Hosotani 
mechanism~\cite{Hosotani:1983xw} (cf.~also~\cite{Candelas:1985en, 
Witten:1985xc}). We emphasize that our interest will be in the quantized, 
non-local realization of gauge symmetry breaking~\cite{Witten:1985xc} (see 
also~\cite{Friedmann:2002ty} and refs. therein) rather 
than in the continuously varying Wilson lines usually associated with the 
term Hosotani-mechanism. String-theoretic implementations following~\cite{ 
Scherk:1979zr} focussed mainly on SUSY breaking~\cite{Rohm:1983aq}. They 
were later realized in explicit orbifold constructions~\cite{Kounnas:1989dk} 
and further developed both in heterotic~\cite{Kiritsis:1997ca} and type I 
model building~\cite{Antoniadis:1999xk}.

It is possible to use the non-local mechanism to break a GUT gauge group to 
the SM at an energy scale different from the string scale, solving, in 
this way, the string scale/GUT scale problem. The main idea is to build 
an orbifold model where the orbifold  group contains non-freely acting 
operators $g^\prime$, preserving the GUT symmetry, and freely acting 
operators $g$, breaking the GUT symmetry \cite{Witten:1985xc}. Of course,
by this orbifolding, an appropriate SUSY breaking to ${\cal N}=1$ in 4d must 
also be realized. 
\begin{center}
\begin{tabular}{|ccccc|}
\hline&&&&\\[-1ex]
& $\{g^\prime\}$  && $\{g\}$ &
\\[-2ex]
\begin{tabular}{c}
$\mathcal{N}=4$\\[1ex] SO(32) or E$_8\times$E$_8$
\end{tabular}&
$\longrightarrow$&
\begin{tabular}{c}
$\mathcal{N}=*$\\[1ex] $\mathcal{G}_{GUT}\times\mathcal{G}$
\end{tabular}&
$\longrightarrow$&
\begin{tabular}{c}
$\mathcal{N}=1$\\[1ex] $\mathcal{G}_{SM}\times\mathcal{G^\prime}$
\end{tabular}\\[3ex]\hline
\end{tabular}
\end{center}

To be more specific, we consider the internal space $T^6=T^2\times T^2\times 
T^2$, with each of the 2-tori parameterized by a complex number $z_i$, $i=1, 
\,2,\,3$. The torus structure emerges after the identifications $z_i\sim 
z_i+(n+m\tau_i)$, $\tau_i$ being the complex structure moduli of the 2-tori
and $n,\,m\,\in\,\mathbb{Z}$.

Now introduce a group $Z_N\times Z^\prime_M$ with generators 
$g$ and $g^\prime $. Let $g$ act as a translation in one of the 6 compact 
dimensions, e.g., 
\beq
g:\mbox{Re}(z_1)\rightarrow\mbox{Re}(z_1)+a\,,
\eeq
while the action in the rest of the space is generic. The full action of $g$ 
is then free and we can embed $g$ in the gauge group in a 
GUT-symmetry-breaking way. Notice that, as discussed earlier, $g$ breaks 
SUSY to $\mathcal{N}\ge 2$ (or to $\mathcal{N}=0$, which is not interesting 
in our context). Thus, we clearly need an extra orbifold operator $g^\prime$ 
generating the second $Z^\prime_M$. We do not require $g'$ to act
freely which means that, to have non-local GUT-breaking, the gauge embedding 
of $g^\prime$ must be GUT-preserving. In order to reduce SUSY to $\mathcal{N} 
=1$ in 4d, it is necessary that $g^\prime$ contains a rotation involving 
Re$(z_1)$. Furthermore, given the non-free action of $g'$, the other 
GUT-breaking operators $g^a\cdot {g^\prime}^b$ ($a,b\in\{0,\cdots, 
N\!-\!1\}$, $a\neq 0$) act, in general, non-freely. This can be avoided by 
a careful choice of the actions of $g$ and $g'$ in the rest of the internal 
dimensions. Remarkably, these actions can not be just standard rotations 
around the origin. 

We now provide two simple examples for $N=M=2$ and for $N=M=3$, leaving a 
complete classification of this type of orbifold models for future work. In 
the first case we choose
\bea
\label{z2freelyacting}
&&\hspace{-1cm}
g:z_1\rightarrow z_1+\frac{1}{2},\hspace{.6cm}
g:z_2\rightarrow -z_2,\hspace{1.6cm}
g:z_3\rightarrow -z_3,\,\,\,\,\, \mbox{GUT-breaking;}\\
&&\hspace{-1.1cm}\label{z2freelyacting2}
g^\prime :z_1\rightarrow -z_1,\hspace{1cm}
g^\prime :z_2\rightarrow -z_2+\frac{1}{2},\hspace{.7cm}
g^\prime :z_3\rightarrow z_3,\,\,\,\,\,\, \mbox{GUT-preserving.}
\eea
Note that the GUT-breaking operator $g$ acts as a rototranslation. It is 
easy to check that the second GUT-breaking operator $g\cdot {g^\prime}$ also 
acts freely (the key being that the action of $g'$ on the second torus is 
not just a rotation around the origin). The breaking is delocalized along 
the real parts of $z_1$ and $z_2$. In the limit where the two corresponding 
radii become large, one recovers the 6d field theory model of~\cite{ 
Hebecker:2003we}. (The underlying non-trivial 2d topology, which is that of 
the projective plane, has been mentioned before in the non-SUSY 
context~\cite{Hall:2001tn}.)

In the second case, since we have $Z_3$-rotations, we need 
$\tau_i=e^{\pi i/3}$. We choose
\bea
&&\hspace{-1cm}
g:z_1\rightarrow z_1+\frac{\tau+1}{3},\hspace{.3cm}
g:z_2\rightarrow e^{2\pi i/3}z_2,\hspace{2.2cm}
g:z_3\rightarrow e^{-2\pi i/3}z_3;\\
&&\hspace{-1.1cm}
g^\prime :z_1\rightarrow  e^{4\pi i/3} z_1,\hspace{.8cm}
g^\prime :z_2\rightarrow  e^{-2\pi i/3} z_2+\frac{\tau+1}{3},\hspace{.33cm}
g^\prime :z_3\rightarrow  e^{-2\pi i/3} z_3+\frac{\tau+1}{3}.
\eea
As before, $g$ is GUT-breaking and $g^\prime$ is GUT-preserving.

It is interesting to study the relation of these kinds of models to 
constructions with gauge symmetry breaking by Wilson lines, both in the 
continuous and quantized case. Models with quantized Wilson lines are 
conceptually close to our discussion since a quantized Wilson line is obtained 
if a freely acting symmetry $h^\prime$ (a translation in the internal 
space) combined with a (quantized) rotation in the gauge bundle is modded 
out. The quantization arises from the consistency with other (rotational) 
symmetries modded out in parallel. As we will demonstrate below, our non-local 
breaking can be realized, quite analogously, by consistently combining Wilson 
lines with orbifold operations. However, in our case the latter can not be 
merely standard rotations. Rather, they must include rototranslations. 

To make the above discussion more explicit, consider the following 
$Z_2\times Z^\prime_2\times Z^{\prime \prime}_2$ example, 
where the various factors of the orbifold group are generated by $g$, 
$g^\prime$ and $g^{\prime\prime}$, 
\bea
&&\hspace{-1cm}
g:z_1\rightarrow z_1+\frac{1}{2}\,,\hspace{.6cm}
g:z_2\rightarrow z_2\,,\hspace{1.6cm}
g:z_3\rightarrow z_3\,,\\
&&\hspace{-1.1cm}
g^\prime :z_1\rightarrow z_1\,,\hspace{1.3cm}
g^\prime :z_2\rightarrow -z_2+\frac{1}{2}\,,\hspace{0.43cm}
g^\prime :z_3\rightarrow -z_3\,,\\
&&\hspace{-1.17cm}
g^{\prime\prime} :z_1\rightarrow -z_1\,,\hspace{.87cm}
g^{\prime\prime} :z_2\rightarrow -z_2\,,\hspace{1.18cm}
g^{\prime\prime} :z_3\rightarrow z_3+\frac{1}{2}\,.
\eea
We choose the actions of $g^\prime$ and $g^{\prime\prime}$ in the gauge bundle 
to be GUT-preserving while that of $g$ is GUT-breaking. Actually, the 
GUT-breaking operator $g$ is nothing else than a quantized Wilson line. 
However, in general this is not sufficient to make the overall GUT breaking 
non-local. All the other GUT-breaking operators, $g\cdot g^\prime$, $g\cdot 
g^{\prime\prime}$, and $g\cdot g^\prime \cdot g^{\prime\prime}$, have to act 
freely as well. For this it is crucial that $g^{\prime\prime}$ (the operator 
rotating the axis along which $g$ is a translation) is not a pure rotation, 
but rather a rototranslation ($g^{\prime\prime} :z_3\rightarrow z_3+ 
\frac{1}{2}$). Furthermore, the actions of $g'$ and $g''$ on $z_2$ are not
both rotations around the same axis. Rather, the choice of axes is such that 
$g'\cdot g''$ is a translation in $z_2$, ensuring that $g\cdot g'\cdot g''$ 
is non-local. 

The relation of non-local quantized gauge symmetry breaking to continuous 
Wilson lines is instead less direct. It is well
known~\cite{Ibanez:1987xa,Mohaupt:1993fb} 
and has already been emphasized above that such Wilson lines are a 
fundamental tool in realistic orbifold model building. Continuous Wilson 
line breaking is, in general, truly non-local since the corresponding 
non-contractible loops can not be shrunk to zero length at an orbifold fixed 
point. Geometrically, they characterize the relative orientation in gauge 
space of breaking-patterns realized at spatially separated fixed points. 
However, a GUT breaking by continuous Wilson lines is, from a field theory 
point of view, nothing but the breaking by a Higgs field developing a VEV 
along a flat direction. This is a major difference with the non-local 
quantized Wilson line case, where the breaking scale is set by the size of 
certain compact radii. Of course, on top of non-local (as well as local)
breaking by quantized Wilson lines, additional breaking by continuous 
Wilson lines may be present. This group-theoretically distinct, extra
breaking is, in  general, rank-reducing.

\subsection{Local breaking and modified running above the\\ compactification 
scale}\label{locbrk}
In this subsection we discuss the implications of a gauge symmetry breaking 
that is local from the point of view of the $4+d$ dimensional field theory; 
we can be very brief since this situation has recently been discussed in some 
detail in the context of orbifold GUTs. We are assuming, as before, 4d MSSM 
running up to a scale $M_c\simeq R_l^{-1}$ set by the $d$ large 
compactification radii. Above that scale, we are facing a $4+d$ dimensional 
field theory. In our context, the simplest models are those where this field 
theory has a unified gauge symmetry in the $4+d$ dimensional bulk, which 
is broken only at certain singular points in the $d$-dimensional compact 
space.\footnote{
In 
addition to the local breaking at singularities, the gauge symmetry 
may, in principle, also be broken locally at every point in the $4+d$ 
dimensional bulk. As discussed in detail in~\cite{Hebecker:2004xx}, the 
corresponding power-like and potentially large threshold corrections will, 
in many cases, be calculable within the framework of low-energy effective 
field theory.
} 
The logarithmic running of the differences of the three inverse SM 
gauge couplings continues, in general, above $M_c$ and stops at either 
a strong-interaction or string scale $M>M_c$. At this scale, the three 
gauge couplings have to be equal up to possible small differences due to 
threshold corrections arising from physics at the scale $M$. 

There are two potential disadvantages of this approach when compared to 
the non-local breaking discussed above. First, the running above $M_c$ is 
model-dependent (and in general different from the MSSM running). It can and 
must be adjusted to ensure unification at the scale $M$. This implies that, 
to the extent to which $\ln(M/M_c)$ is not negligible compared to 
$\ln(M/m_Z)$, the purely MSSM-based prediction of gauge unification is lost. 
Second, as already discussed in Sect.~\ref{fac} (see also further down), the 
UV completion relevant at the scale $M$ (especially in the vicinity of the 
orbifold fixed points) will, in general, not be perturbative. Thus, in many
otherwise attractive models, threshold corrections may be uncalculable in
principle, and no tests with an accuracy better than the familiar one-loop
MSSM running may be possible. Of course, the advantage of this local-breaking 
scenario is that any known heterotic orbifold model can be deformed by making 
some of its radii large, providing many potentially interesting and 
realistic examples. 

To be more quantitative, we have to make use of the detailed results of 
Sects.~\ref{lr} and~\ref{fac}. In the most promising case of $d=1$, we are 
dealing, above the scale $M_c$, with a 5d field theory compactified on an 
interval (which can be viewed as an $S^1/Z_2$ or $S^1/(Z_2\times Z_2')\,$). 
If $M_c>3.4\times 10^{16}$ GeV, the heterotic UV completion can be realized 
in the perturbative regime. This situation may look somewhat contrived since 
the gauge couplings meet at $M_{GUT}$ by pure chance, continue to run in an
MSSM-like way up to $M_c$, and finally meet again at $M\simeq m_H$ after 
the modified logarithmic running between $M_c$ and $M$. However, there 
is nothing wrong with this option in principle. 

If $M_c<3.4\times 10^{16}$ GeV (including in particular the case where 
$M_c<M_{GUT}$), the heterotic theory at scale $m_H$ is strongly coupled, and 
dangerous non-perturbative light states may be present. The analysis of
Sect.~\ref{fac} has shown that no such states appear below a scale
\beq
M\simeq 50M_c\,(M_c/3.4\times 10^{16}\,\,\,\mbox{GeV})^{2/5}\,.\label{uv1}
\eeq
We are unable to discuss the presumably non-perturbative physics near the 
orbifold fixed points at energies $\sim M$. However, it is natural that any 
brane-localized corrections to gauge unification arising from this type of 
physics will be suppressed by a volume factor $M_c/M$, which is comfortably 
large for $M_c\sim M_{GUT}$. (We assume that the branes have an effective 
thickness $\sim M^{-1}$.) However, we also see that, when $M_c$ is lowered 
significantly below the GUT scale, bulk dominance (and hence 
calculability of gauge coupling unification) is gradually lost. 

As a side remark, we note that the lowest value of $M_c$ (for $d=1$) 
compatible with the heterotic perturbative regime, $M_c\simeq m_H/50
\simeq 3.4\times 10^{16}$ GeV, is intriguingly close to $M_{GUT}$. 
To make quantitative use of this fact, we would need to find a model where 
the GUT symmetry is broken at the boundary of an interval of size 
$M_c^{-1}$ and no differential running of (brane localized contributions to) 
SM gauge couplings occurs above that scale. 

For brane-localized GUT breaking in models with $d\ge 2$ large radii, the 
quantitative situation with respect to gauge coupling unification appears 
to be more challenging. In particular, as discussed in detail in the $d=2$ 
case in Sect.~\ref{twolarge}, there are massive string theory states very 
close to 
the compactification scale $R_l^{-1}$. In the non-local breaking case, 
where their effects are exponentially suppressed, we could hope that a
small hierarchy, say $M/M_c\sim MR_l\sim 3$, is sufficient. In the 
local breaking case with an effective brane thickness $\sim M^{-1}$, the 
volume suppression of localized GUT breaking effects by $(MR_l)^{-d}\sim
1/9$ is dangerously weak. Of course, we can hope that specific models have
smaller non-perturbative corrections, e.g. because of the `Wilson line 
improvement' discussed earlier.

\section{Some Examples}\label{ex}
\subsection{A $Z_2\times Z^\prime_2$ model with non-local gauge symmetry 
breaking}\label{ex1}
In Sect.~\ref{twolarge} we showed that highly anisotropic 
compactifications of the SO(32) heterotic string with $d\!=\!2$ large 
radii $R_l\simeq (2\times 10^{16}\,\mbox{GeV})^{-1}$ are viable. 
Subsequently, we explained in Sect.~\ref{nln} how a $Z_2\times 
Z^\prime_2$ orbifold model with non-local gauge symmetry breaking can be 
built, realizing a breaking scale proportional to the compactification 
scale. Combining these possibilities, a realistic string model with a 
unified gauge group broken at $M_{GUT}\sim 2\times 10^{16}\,{\rm GeV}$ 
can, in principle, be constructed. In the present section we present a
model which, though not complete, can form the basis of a realistic model 
implementing the above ideas. 

\begin{figure}[t]
\begin{center}
\hspace{0cm}
\includegraphics[scale=0.53]{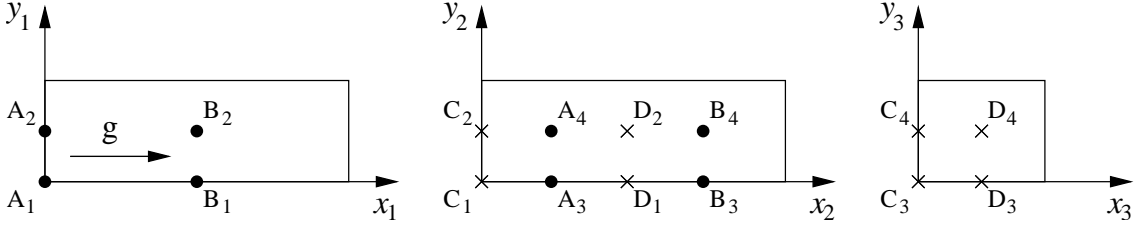}\vspace{0cm}
\caption{\small \it The structure of the internal space $(T^2\times T^2\times 
T^2)/(Z_2\times Z^\prime_2)$. Note that the two larger dimensions are 
shared between the first and the second torus. The action of $Z^\prime_2$ 
is a rotation in the first two tori with fixed points shown as black dots 
with labels $A$ and $B$. The action of $Z_2$ is instead a translation along 
$x_1$ (arrow) and a rotation in the second and third torus with would-be 
fixed points shown as crosses with labels $C$ and $D$.
}\label{toriZ2}
\end{center}
\end{figure}

We parameterize the internal space as in Sect.~\ref{nln}, but with 
dimensionful parameters $z_i=x_i+i\,y_i$, so that the periodicities are 
$z_i\sim z_i+2\pi R_i(m+n\tau_i)$. For simplicity, we consider rectangular 
tori with 6 radii $R_i$ and $R_{3+i}$ (for $i=1,2,3$) where $\tau_i= 
i\, R_{3+i}/R_{i}$ is purely imaginary. The geometric action of 
Eqs.~(\ref{z2freelyacting}) and (\ref{z2freelyacting2}) is then rewritten as
\bea
&&g:z_1\rightarrow z_1+\pi R_1,\,\,\,\,\,\,\,\hspace{0.5cm}g:z_2
	\rightarrow -z_2,\hspace{2.0cm}
g:z_3\rightarrow -z_3,\\
&&g^\prime:z_1\rightarrow -z_1, \hspace{1.7cm} g^\prime:z_2
	\rightarrow -z_2+\pi R_2,
\,\,\,\hspace{.5cm}g^\prime:z_3\rightarrow z_3,
\eea
as summarized in Fig.~\ref{toriZ2}.

The GUT group is part of the gauge group left unbroken by the action of 
$g^\prime$ in the gauge bundle as well as by possible Wilson lines. Allowed 
choices for this group are discussed below.\footnote{
For 
the moment, the crucial point in this identification is the breaking scale, 
which is high (string scale) for SO(32)$\,\rightarrow\mathcal{G}_{GUT}\times 
\mathcal{G}$ and low ($M_{GUT}$) for $\mathcal{G}_{GUT}\times \mathcal{G} 
\rightarrow\mathcal{G}_{SM}\times \mathcal{G}^\prime$.
}
The GUT group is then broken by $g$. Since this operator includes a 
translation along $x_1$, it has no fixed points in the internal space and 
its gauge symmetry breaking scale is $\sim R_1^{-1}$. The only 
other GUT-symmetry-breaking operator is $g\cdot g^\prime$, which includes 
a translation along $x_2$ and correspondingly has a GUT symmetry breaking 
scale $\sim R_2^{-1}$. We can then take $R_1\simeq R_2\simeq (2\times 
10^{16}\,{\rm GeV})^{-1}$ and $R_i\simeq \alpha_H'^{1/2}\simeq (8.6\times 
10^{17}\,{\rm GeV})^{-1}$ for $i=3,\cdots,6$. Recall that, as explained in 
detail in Sect.~\ref{twolarge}, potentially dangerous light states of the 
6d effective field theory are somewhat closer to $M_{GUT}$ than the fairly 
high scale $\alpha_H'^{-1/2}$. 

\begin{figure}[t]
\begin{center}
\includegraphics[scale=0.45]{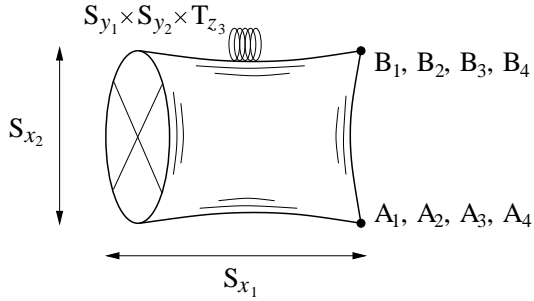}\vspace{-.3cm}
\caption{\small \it Visualisation of the 6d internal space after orbifolding. 
The two larger extra dimensions $S_{x_1}$ and $S_{x_2}$ parameterize a 
`half-pillow' with crosscap identification on the left. Away from the two 
conical singularities on the right, every point of this 2d space supports 4 
smaller extra dimensions with $T^4$ geometry.
}\label{twodimspace}
\end{center}
\end{figure}

It is also interesting to note how, scanning energy scales in effective 
field theory, one first finds the 2d compact space of~\cite{Hebecker:2003we}, 
parameterized here by the coordinates $x_1$ and $x_2$. Then, at higher 
energies, each of the points of this 2d space (away from the two orbifold 
singularities) can be resolved as a $T^4$, parameterized by $y_1$, $y_2$ and 
$z_3$ (see Fig.~\ref{twodimspace}). 

\subsubsection{Gauge embedding: a toy model}
We now explicitly discuss the gauge embedding of the actions of $g$ and 
$g^\prime$. We begin with a simple example and continue, in the next 
subsection, with the much wider class of models with quantized Wilson 
lines including, in particular, an SO(10) GUT model broken non-locally to 
SU(4)$\times$SU(2)$\times$SU(2). 

The most general gauge embedding of the action of $g'$ and $g$ can be 
characterized by shift vectors
\beq
V^\prime=\frac{1}{2}\left(1^n,\,0^{16-n}\right),\,\,\,\,
V=\frac{1}{2}\left(0^p,1^m,\,0^{16-m-p}\right)\,\label{vees}
\eeq
with $p<n<p+m$. The familiar requirements of modular invariance (for a 
recent collection of the relevant formulae see, e.g.,~\cite{Forste:2004ie}) 
impose $n=2\,{\rm mod}\,4$, $m=2\,{\rm mod}\,4$, and $p$ odd. The gauge 
symmetry breaking by $Z^\prime_2$ (at the high scale) is
\beq
SO(32)
\begin{tabular}{c}
local\\[-1ex]
$\longrightarrow $
\end{tabular}
SO(2 n)\times SO(32-2n),
\eeq
where $n=2,\,6,\,10,\,14$ are possible.\footnote{
There
is an obvious duality between $n$ and $16-n$, but we find it convenient not 
to eliminate the resulting double counting.
}
The breaking by $Z_2$ (at the GUT scale) is
\bea
&&SO(2n)
\begin{tabular}{c}
non local\\[-1ex]
$\longrightarrow $
\end{tabular}
SO(2 p)\times SO(2n-2p),\\
&&SO(32-2n)
\begin{tabular}{c}
non local\\[-1ex]
$\longrightarrow $
\end{tabular}
SO(2m+2p-2n)\times SO(32-2m-2p).
\eea
Within this toy model we can identify, for example, `$\mathcal{G}_{GUT}$'
$\equiv\,\,$SO$(2n)$ and `$\mathcal{G}_{SM}$'$\equiv\,\,$SO$(2 p)\times$SO$(
2n-2p)$. Clearly, one is severely limited by the fact that the GUT group can 
only be SO(12), SO(20) or SO(28).

\subsubsection{A Pati-Salam model with two Wilson lines}
The situation can be greatly improved by adding two Wilson lines to the
system. In the next example, the $Z^\prime_2$ projection in the presence of 
Wilson lines breaks, at the high scale, SO(32) to SO(10)$\times\mathcal{G}$. 
We can now identify the GUT group with SO(10) and realize its further 
breaking to the Pati-Salam group SO(6)$\times$SO(4)$\,\equiv\,$SU(4)$\times
$SU(2)$\times$SU(2) by the free action of $g^\prime$. Explicitly, we take
\beq
V^\prime=\frac{1}{2}\left(1^{10},0^6\right),\,\,\,\,
V=\frac{1}{2}\left(0^3,1^2,0^4,1^4,0^3\right).
\eeq
Note that these vectors are just a specific case of Eq.~(\ref{vees}), where 
$n=10$, $m=6$, $p=7$ and Cartan generators have been reshuffled to bring $V$ 
to a form which will be more convenient later on. We also introduce the 
Wilson lines 
\beq
A_1=\frac{1}{2}\left(0^6,1^8,0^2\right),\,\,\,\,
A_2=\frac{1}{2}\left(0^5,1,0^3,1^2,0^4,1\right)
\eeq
in $y_1$ and $y_2$ direction. Since the corresponding radii $R_4$ and $R_5$ 
are very short, the breaking introduced by these two Wilson lines is a 
high-scale breaking, realized everywhere in the 6d bulk spanned by the 
large radii $R_1$ and $R_2$ and the 4d Minkowski space.  It is easy to see 
that the combined high-scale gauge symmetry breaking due to $V^\prime$, 
$A_1$ and $A_2$ is
\bea
{\rm SO}(32)
\a
\begin{tabular}{c}
$V^\prime$\\[-1ex]
$\longrightarrow $
\end{tabular}\a
{\rm SO}(20)\times {\rm SO}(12)
\begin{tabular}{c}
$A_1$\\[-1ex]
$\longrightarrow $
\end{tabular}
{\rm SO}(12)\times {\rm SO}(8)\times {\rm SO}(8)\times {\rm SO}(4)\nonumber\\
\a
\begin{tabular}{c}
$A_2$\\[-1ex]
$\longrightarrow $
\end{tabular}\a
{\rm SO}(10)\times {\rm SO}(2)\times {\rm SO}(6)\times {\rm SO}(2)^2\times 
{\rm SO}(6)\times {\rm SO}(2)^2\,.
\eea
Identifying the GUT group with SO(10), the further breaking by $g$ 
(characterized by $V$) can be recognized as the desired non-local GUT breaking 
SO(10)$\,\rightarrow\,$SO(6)$\times$SO(4). At this point, we have to assume 
that a further breaking to the SM gauge group can be realized either 
field-theoretically (at least the doublet-triplet splitting problem is now 
solved) or in a better string model. We leave the search for such a more 
complete model for the future. 

The choice of compactification radii shown in Figs.~\ref{toriZ2} 
and~\ref{twodimspace} and discussed in Sect.~\ref{twolarge} ensures 
that the GUT symmetry breaking has the correct scale, couplings unify 
quantitatively in the perturbative domain of 6d field-theory, and that the 
perturbative calculation of the string spectrum can be trusted qualitatively. 
Furthermore, we assume bulk dominance (which should, according to 
Sect.~\ref{twodimspace}, be marginally valid) and recall that we have 
${\cal N}=4$ SUSY away from the fixed points. Under these two premisses, 
the SUSY-protected relations between $M_P$, $\alpha_{GUT}$, $g_H$ and 
$\al_H$ can be trusted quantitatively even though string-theoretic 
perturbativity at the high scale is, in general, lost.

\subsection{Orbifold GUTs in 5 dimensions}\label{OGUT5D}
This section is devoted to the implications of our general string-theoretic 
perturbativity analysis (and in particular the results of Sects.~\ref{fac}
and~\ref{locbrk}) for specific 5d unified models with brane localized GUT
breaking (orbifold GUTs). 

The simplest realistic orbifold GUTs have SU(5) symmetry in 5d, broken to 
the standard model group by appropriate $Z_2$ boundary conditions at one 
of the two boundaries (SM brane)~\cite{Kawamura:2000ev,Hall:2001pg,
Hebecker:2001wq}. In a minimalist setting, the two Higgs doublet superfields 
are introduced at this SM brane~\cite{Hebecker:2001wq}. (At our level of 
precision, the bulk or brane nature of SM matter is irrelevant because it 
comes in full SU(5) multiplets.) In the setting with brane-localized Higgs 
doublets, the relative running of the differences of inverse gauge couplings
(differential running) above $M_c$ is very close to the 4d MSSM: The 
relevant $\beta$ function coefficients are $(8/5,-2,-5) 
$~\cite{Hebecker:2001wq}, which have to be compared with the MSSM values 
$(33/5,1,-3)$. To quantify the similarity of these two sets of coefficients, 
consider ratios of the differences $\Delta\alpha_{ij}=\alpha_i^{-1}- 
\alpha_j^{-1}$ created by the running ($i,j$ refer to the three SM 
gauge group factors), for example
\beq
\left(\frac{\Delta\alpha_{12}}{\Delta\alpha_{23}}\right)_{\mbox{MSSM}}=1.4
\qquad\mbox{and}\qquad\left(\frac{\Delta\alpha_{12}}{\Delta\alpha_{23}}
\right)_{\mbox{minimal orbif. GUT above $M_c$}}=1.2\,. 
\eeq
The similarity of these two ratios implies that unification will not be 
drastically affected even if part of the usual MSSM running is replaced by 
the modified running above $M_c$. The precision loss in the unification of 
the $\alpha_i^{-1}$ can be quantified as 
\beq
\Delta\alpha_{\rm err}\simeq \frac{\ln(M_{GUT}/M_c)}{\ln(M_{GUT}/m_Z)}\cdot 
\frac{|1.2-1.4|}{1.4}\cdot 30\,, 
\eeq
based on the total change of the difference of, e.g.,  the first two inverse 
couplings in conventional MSSM unification, $\Delta\alpha_{12}(m_Z)\simeq 
30$. 

However, even though the modified running does not significantly affect 
unification precision, it does change the final unification scale because 
the `speed' of the modified running is reduced. Estimating the reduction 
factor as $(8/5\!+\!2)/(33/5\!-\!1)\simeq (-\!2\!+\!5)/(1\!+\!3)\simeq 0.7$, 
this raised final unification scale (which we identify with the UV scale $M$ 
of the relevant model) is determined by 
\beq 
\ln(M_{GUT}/m_Z)\simeq \ln(M_c/m_Z)+0.7\,\ln(M/M_c)\,.\label{uv2}
\eeq

Now recall that in orbifold GUTs as well as in corresponding string models
with strong coupling at $M$, a limit on the precision of gauge unification 
is set by the ratio $M/M_c$. If, on the one hand, $M_c$ is large (near 
$M_{GUT}$), Eq.~(\ref{uv2}) fixes the ultimate unification scale $M$ 
dangerously close to $M_c$. If, on the other hand, $M_c$ is small, 
Eq.~(\ref{uv1}) suggests that the UV scale $M$ (and thus $M/M_c$) is low 
because of the presence of light non-perturbative states. The optimal value 
is $M_c\simeq 2.6\times 10^{15}$ GeV, with $M/M_c \simeq 20$. Although not 
comfortably large, this value may still be consistent with the basic orbifold 
GUT idea of gauge unification based on bulk dominance. 

We expect the following features arising in the above specific model to be 
generic: If the differential running above $M_c$ is similar to the MSSM 
running but `slower', then $M>M_{GUT}>M_c$ and a certain validity range of 5d 
field theory (ensuring precision gauge coupling unification even in the case 
of a non-perturbative stringy UV completion at $M$) can be realized. The 
situation becomes better the slower the differential running above $M_c$ 
is. Ideally, one would want a model where, in spite of hard (from the 
field-theoretic perspective) GUT breaking at the boundary, no differential 
running above $M_c$ occurs. This would clearly require a very special bulk 
and brane field content.

In this paper, we have restricted ourselves to simple leading-log running, 
ignoring all other loop corrections. From this perspective, MSSM unification 
is perfect within the expected 
accuracy\footnote{ 
For example, adopting the values of $\alpha_{1,2}(m_Z)$ given in Eq.~(6.22) 
of~\cite{Fusaoka:1998vc} and identifying the effective SUSY breaking scale
with $m_Z$, one-loop running predicts $M_{GUT}\simeq 2\times 10^{16}$ GeV 
(the value used above) as well as an acceptable strong coupling $\alpha_3
(m_Z)\simeq 0.114$.
}
and the problem is merely to achieve consistency with string theory and the 
scale of gravity. A more careful analysis, including loop corrections at the 
weak/SUSY-breaking scale and 2-loop running, reveals that the $\alpha_3$ 
value predicted by grand unification is somewhat high (see~\cite{
Eidelman:2004wy} and refs. therein). In our opinion, this is not particularly 
serious in view of the significant high-scale uncertainties present, in 
particular, in string-based orbifold GUT models (recall the brane-localized 
non-universal gauge-kinetic terms). However, taking the discrepancy seriously,
it was suggested early on that one can use the deviation of the differential 
running above $M_c$ from the MSSM running to account more accurately 
for the measured value of the strong coupling~\cite{Hall:2001xb}. Without 
going into detail, we emphasize that the above string-theory-based 
restrictions on the size of the ratio $M/M_c$ apply and it may be necessary 
to reconsider such orbifold GUT models (with `improved' unification) in this 
context. 

We discussed 5d orbifold GUTs mainly using the perturbativity 
analysis of the SO(32) heterotic string. The E$_8\times$E$_8$ case is also 
very relevant even though, as explained in Sect.~\ref{radiiE8}, this scenario 
appears to be more constrained. A recent search for phenomenologically 
appealing string-theoretic orbifold GUT models of this type appeared in~\cite{
Kobayashi:2004ud}, where the unification scale problem was also discussed. 
The authors conclude that one is driven into the $M$ theory regime with 
the scales $\rho$ and $R_l$ very close (although the problem of small $R_s$ 
is not analysed). This is consistent with our findings, which we interpret 
more strongly as implying a serious problem for precision gauge coupling 
unification (insufficient volume suppression of non-perturbative brane 
effects). 

We finally note that, in our understanding, the discussion of local breaking 
in this subsection is close in spirit to the very recent analysis 
of~\cite{Ross:2004mi}. There, a mildly increased radius is used to bring the 
unification scale closer to the string scale. We believe that the 
underlying effect is related to the slower differential running above $M_c$ 
used in this subsection and discussed earlier in the context of orbifold 
GUTs.

\section{Conclusions}\label{conc}
We have discussed mass scales in heterotic string model building, attempting 
to use the compactification radii to explain the well-known discrepancy 
between the SUSY unification scale $M_{GUT}$ and the string scale 
$M_{string}$. Our starting point was the careful definition of a proper 
string coupling constant $g$ such that $g\!=\!1$ characterizes quantitatively 
the boundary of the perturbative regime of the heterotic string. We found 
surprisingly good agreement between a duality based and an effective field 
theory based definition of this string coupling.

Having rederived, with the above perturbativity definition, the well-known 
problems of isotropic compactifications, we focussed our attention on highly 
anisotropic orbifold geometries. Taking one of the compact radii larger than 
the other four, it is easy to move the corresponding compactification scale 
$M_c$ to the GUT scale without leaving the perturbative regime of string 
theory. However, it is unfortunately impossible to identify $M_c$ with the 
GUT breaking scale. The basic reason is that the compact geometry is that 
of an interval and that the GUT group is, in general, broken at one or both 
of its boundaries. The logarithmic running of gauge coupling differences 
then does not stop at $M_c$ but continues up to the string scale. 

Nevertheless, this effective 5d scenario is interesting since the modified
differential running between $M_c$ and the string scale may be used
to realize a precise gauge coupling unification in spite of the fact that 
the running does not stop at $M_{GUT}$. In fact, lowering the smaller radii 
to below the string length and/or entering the string-theoretic strong 
coupling regime, it is even possible to take $M_c$ significantly below 
$M_{GUT}$. For example, we find it possible to have $M_c\simeq 2.6\times 
10^{15}$ GeV while losing perturbative control only at a scale $M\sim 20 
M_c$. In this situation, the bulk suppression factor $1/20$ ensures 
(moderately) precise unification in spite of non-perturbative brane-localized 
effects at the scale $M$. Furthermore, the somewhat slower differential 
running above $M_c$ of the simplest orbifold GUTs improves the unification 
precision at the slightly high scale $M>M_{GUT}$. 

An obvious disadvantage of the above and other related 5d scenarios is the 
model dependence introduced by the modified differential running above $M_c$. 
By contrast, in the case of two larger radii, it is possible to construct 
models where precise unification occurs just on the basis of the MSSM 
running between the electroweak scale and $M_{GUT}$. 

The key is the observation that, even in the case of simple toroidal 
orbifolds, the geometry of the two larger extra dimensions of these 6d 
models can be sufficiently complicated to allow for a non-local GUT breaking 
(with the scale set by $M_c\simeq M_{GUT}$) together with an appropriate 
reduction of SUSY to ${\cal N}\!=\!1$. One has to pay the price of entering 
the non-perturbative regime of the underlying string theory sufficiently 
deeply so that one has to expect a breakdown of the 6d field theory very
close to $M_c$. However, since the GUT breaking is non-local, the resulting 
non-perturbative contributions to gauge coupling differences are 
exponentially suppressed and a small hierarchy between $M_c$ and the strong 
coupling scale $M$ may be sufficient. 

The technical tool used in the construction of the above models with two 
large radii and non-local breaking is the freely acting orbifold. 
The gauge symmetry is broken by non-local discrete 
Wilson lines, which can be viewed as an extension of the widely used method 
of discrete Wilson line breaking. 
To ensure that such Wilson lines introduce a non-local 
breaking (i.e., there are no fixed points with reduced gauge symmetry) it 
is necessary, starting from a torus with Wilson lines, to mod out 
rototranslations rather than simple rotations. We provide an explicit 
example where the original SO(32) gauge symmetry is broken near the string 
scale to a 6d SO(10) and then, by a non-local breaking at the GUT scale, to 
an SU(4)$\times$SU(2)$\times$SU(2) model in 4d. 

In analysing how small the smaller of the orbifold radii can be made and 
at which energy scale string-theoretic perturbativity is lost, we had to rely
heavily on $S$ and $T$ dualities. In particular, the above statements about 
the perturbativity loss in 5d or 6d field theory were based on going, via a
network of dualities, to a weakly coupled description and identifying the 
lowest-lying (and thus most dangerous) massive string state. Our calculations 
were conservative in that we used toroidal compactifications and $T$ dualities 
in their basic form. However, we know that in the presence of Wilson lines
(which are in any case necessary in realistic model building) $T$ duality 
constraints on the smallness of the small radii will be relaxed. Thus, we 
are confident that a large class of interesting models with the above 
promising features with respect to gauge coupling unification exist. 

The $T$ dualities mentioned above and used extensively in our analysis of 
the type I dual of the SO(32) heterotic string are not available for the $M$ 
theory dual of the E$_8\times$E$_8$ heterotic theory. Thus, our ability to 
analyse this case is significantly weaker (except for the special case with 
Wilson line breaking to SO(16)$\times$SO(16) which is dual to the SO(32) 
theory). Using conservative estimates on the minimal allowed size of the 
small compact dimensions, we find that both in the effective 5d and 6d case 
generic E$_8\times$E$_8$ string theory completions are more constrained than 
the corresponding SO(32) models. 

We believe to have shown that the GUT scale/string scale problem points to
string compactifications where one or two of the compact radii are 
exceptionally large. The related phenomenological requirements, which have 
to a certain extent been explored in the field theory context, leave a wide 
field of possible activities on the side of string model building.
We are particularly intrigued by the unexplored class of orbifold models
where non-local discrete Wilson lines break the GUT group at the
compactification scale.

\noindent
{\bf Acknowledgements}: We would like to thank Ignatios Antoniadis, 
Wilfried Buchm\"uller, Jan Louis, Stefan Groot Nibbelink, Marco Serone, 
Stephan Stieberger, Stuart Raby, Michael Ratz and Rodolfo Russo for useful
comments and discussions.

\end{document}